\documentclass[12pt]{article}
\usepackage{cite}

\begin{document}


\begin{center}
{\Large \bf Models of fragmentation with power law log-normal distributions}
\vskip 0.6cm

{\large Z.~Tavassoli \footnote{corresponding author: zohreh.tavassoli@brunel.ac.uk}} \\

\vskip 0.3cm
{\it Department of Mathematical Sciences, Brunel University,\\
Uxbridge, Middlesex, UB8 3PH, U.~K.}

\vskip 0.3cm
{\large and} \\

\vskip 0.3cm
{\large A.~Esmaeilnia Shirvani}\\

\vskip 0.3cm
{\it Physics Department, Shahid Beheshti University, Ewin, Tehran 19834, Iran}
\end{center}

\vskip 0.3cm 
{\noindent \large \bf Abstract}

\vskip 0.3cm
\noindent%
Two models of binary fragmentation are introduced in which a time dependent transition size produces two regions of fragment sizes above and below the transition size. In the models we consider a fixed rate of fragmentation for the largest fragment and two different rates of fragmentation for the two regions of sizes above and below the transition size. The models are solved exactly in the long time limit to reveal stable time-invariant solutions for the fragment size distributions. A rate of fragmentation proportional to the inverse of fragment size in the smaller size region produces a power law distribution in that region. A rate of fragmentation combined of two terms, one proportional to the inverse of the fragment size and the other proportional to a logarithmic function of the fragment size, in the larger size region produces a log-normal distribution in that region. Special cases of the models with no fragmentation for the smaller fragments are also considered. The similarities between the stable distributions in our models and power law log-normal distributions from experimental work on shock fragmentation of long thin glass rods and rupture of mercury droplets are investigated.





\newpage
\section{Introduction}
\label{flpsection1}

Fragmentation is an irreversible kinetic phenomenon which occurs in many physical and chemical processes. Because of broad range of applications, many recent studies have been carried out to investigate the kinetics of the processes by introducing simple fragmentation models \cite{flpref1,flpref2,flpref3,flpref4,flpref5,flpref6,flpref7,flpref8,flpref9,flpref10}. In \cite{flpref1,flpref2,flpref3} Ziff and MaGrady presented a model of fragmentation in which the rate of break up depends on the size of the fragments. A general discussion of the kinetics of continuous fragmentation processes was given by Cheng and Redner in \cite{flpref5}. In \cite{flpref7,flpref8} Krapivsky and Ben-Naim studied the kinetics of random fragmentation of multidimensional objects. In \cite{flpref9} three models of binary fragmentation were investigated analytically in which at each time step, the largest fragment in the system is broken with some externally tuneable probability. In the second model of that paper, at each time step, there is a fixed rate of fragmentation for the largest fragment and a rate proportional to the inverse of the fragment size for all other fragments smaller than the largest one. The model was solved exactly in the long time limit to reveal a power law distribution with an exponent, which depends on the precise details of the fragmentation process.

Surprisingly, there is a similarity between the result of the stable distribution from the second model of \cite{flpref9}, with some experimental and analytical results of fragment size and mass distributions with power law forms in the shock fragmentation \cite{flpref11,flpref12,flpref13,flpref14,flpref15,flpref16,flpref17,flpref18,flpref19,flpref20}. In \cite{flpref10} a generalization of the second model of \cite{flpref9}  was carried out. Some models of binary fragmentation were introduced which revealed composite power law distributions for the mass and size of the fragments \cite{flpref10}.

Another behaviour of distribution which is observed in many different processes has a log-normal form \cite{flpref18,flpref19,flpref20,flpref21,flpref22,flpref23,flpref24}. In \cite{flpref22} a comprehensive study in the log-normal distribution is given. The application of these distributions in many different areas of science is discussed in \cite{flpref22}, such as economics, biology, astronomy, philology, small particle statistics and physical and industrial processes. The log-normal behaviour is known to be able to describe the size distribution well in a wide variety of geological situations \cite{flpref24}, such as that of rocks in a boulder field. In \cite{flpref18} an analytical and numerical study was carried out on the fragmentation of long thin glass rods. Some stochastic models for one-dimensional brittle fracture were introduced. The models successively extended to describe cascade fracture of long thin glass rods. They are found to give the fragment size distribution close to a log-normal one. In terms of rupture in liquids, experiments in rupture of mercury drops were performed \cite{flpref19}. The result of size distribution for the cumulative number of droplets showed a log-normal distribution at the small falling heights. A transition of the distribution from log-normal to scaling behaviour was observed as the falling height was increased. In \cite{flpref20} an experimental work on the shock fragmentation of long thin glass rods was carried out by Ishii and Matsushita. The results of fragment size and mass distributions at small falling heights showed a log-normal form for larger fragments and a power law form for smaller fragments. The crossover was seen to be at length scales around the rod diameter. The mass and size distributions for the larger fragments showed a power law form as the falling height was increased.

This paper is organised as follows. In sections \ref{flpsection2} and \ref{flpsection3} we introduce two new models, which are generalisations of the second model of \cite{flpref9} and the models of fragmentation in \cite{flpref10}. In the models we introduce a time dependent transition size which produces two regions of fragment sizes with different rates of fragmentation above and below the transition size. A rate of fragmentation proportional to the inverse of the fragment size in the smaller size region produces a power law distribution in that region. A rate of fragmentation combined of two terms, one proportional to the inverse of the fragment size and the other proportional to a logarithmic function of the fragment size yield a log-normal distribution in the larger size region. Finally in section \ref{flpsection4} we summarise our conclusions and discuss the application of our results to experimental work on the shock fragmentation of long thin glass rods \cite{flpref20} and the rupture of mercury drops \cite{flpref19}. 
    
For the rest of this introduction we give a very brief summary of previous work on models of binary fragmentation. In these models, the density of fragments of size $y$ at time $t$, $n(y,t)$, evolves according to

\begin{equation} \label{flpeq1}
n(y, t \,+ \,\delta t) = n(y,t) \,- \,\delta t \,n(y,t) \int_0 ^y R(z,y-z)dz \,+ \, 2 \,\delta t \int_y ^{L(t)} R(y,z-y)\,n(z,t)dz,
\end{equation}

\noindent%
where $R(y,z)$ is the intrinsic rate that a fragment of size $(y + z)$ breaks into fragments of size $y$ and $z$. The first term on the right hand side is the contribution from those fragments not chosen for the fragmentation in time $(t, \,t \,+ \, \delta t)$. The second term represents the decrease in the number of fragments of size $y$ by the fragmentation into fragments of size $z$ and $(y - z) (y > z)$. The third term represents the increase in the number of particles of size $y$ due to the fragmentation of particles of size $z (>y)$, such that one of the products is of size $y$.

The upper limit of the second integral on the right-hand side is usually set to some fixed value greater than the size of the largest particle in the system. This value is usually $1$ or $\infty $ depending on the details of the model. To connect with the analysis of our models in the following sections, we have set the upper limit to $L(t)$, the size of the largest particle at time $t$. As $n(y,t) = 0$ for all $y > L(t)$, this choice has no effect on our results.
       
We now turn to a detailed investigation of our models.

\section{Model A}
\label{flpsection2}

Firstly, we introduce a time dependent transition size, $y_m(t)$, between zero and the size of the largest particle in the system at time $t$, $L(t)$. The transition size produces two regions of fragment sizes with different rates of fragmentation above and below the transition size. At each time step there is a fixed probability, $p_1$, for the fragmentation of the largest particle in the system. A particle of size smaller than the transition size breaks with a probability $p_2$ and a rate proportional to the inverse of the fragment size, at each time step. Any other particle of size larger than the transition size can be fragmented, at each time step, either with a probability $p_3$ and a rate proportional to the inverse of the fragment size, or with a probability $p_4$ and a rate proportional to $\displaystyle\frac{\delta t}{y(t)}\ln\left(\frac{y(t)}{\tilde y(t)}\right)$. Here $y(t)$ is the size of the fragment at time $t$ and $\tilde y(t)$ is a size for which the logarithmic rate is zero at time $t$. In the model we have $p_1 + p_2 + p_3 + p_4 = 1$. We also assume that $y_m(t) \geq \tilde y(t)$, such that for all fragments of sizes larger than the transition size, the logarithmic term in the rate of fragmentation is positive. In order to make the model tractable, we will later choose to have $y_m(t)$ proportional to $L(t)$.

In the model the distribution function of daughter fragments is uniform. Therefore the rate of breaking of a particle into two smaller pieces is independent of the daughter fragment sizes. With probability $p_2$ every particle of size smaller than $y_m(t)$ is equally likely to be chosen for the fragmentation. With probabilities $p_3$ and $p_4$, every particle of size larger than the transition size is equally likely to be selected for the fragmentation with the correspondence rate. We assume that in time $(t, t + \delta t)$ the largest size changes from $L(t)$ to $L(t) - p_1\delta L$ and the transition size changes from $y_m(t)$ to $y_m(t) - \delta y_m$.

Then, the fragmentation rate for this model is given by
  
\begin{eqnarray}  \label{flpeq2}
\delta t \,R(y,z)& \,= &\,p_1 \,\frac{\delta L}{L(t)} \,\delta (y\,+\,z-L(t)) \,+ \,p_2 \,\frac{\delta t}{y+z} \,\theta (y_m(t)-(y+z)) \nonumber\\
&  &+\:\left[p_3\,\frac{\delta t}{y+z}\,+\,p_4\,\frac{\delta t}{y+z}\,\ln\left(\frac{y+z}{\tilde y(t)}\right)\right]\theta ((y+z)-y_m(t)). \nonumber\\
\end{eqnarray}	   

\noindent%
The $\delta$-function in the first term on the right-hand side ensures that only the largest particle is fragmented and $\displaystyle\frac{\delta L}{L(t)}$ is the probability of placing the cut in the largest particle in a particular infinitesimal length $\delta L$. The second term on the right-hand side of (\ref{flpeq2}) represents the rate of fragmentation for the particles of sizes smaller than $y_m(t)$ in time $(t, t + \delta t)$. The remaining terms on the right hand side of (\ref{flpeq2}) represent the rate of fragmentation for the fragments of sizes between $y_m(t)$ and $L(t)$ in time $(t, t + \delta t)$. Inserting the rate (\ref{flpeq2}) into the kinetic equation (\ref{flpeq1}) for the fragments in both size regions smaller and larger than $y_m(t)$, yields

\begin{eqnarray}  \label{flpeq3}
n_1(y,t+\delta t)& = &n_1(y,t)\;-\;p_2\,\delta t \,n_1(y,t)\;+\;2\,p_1 \,\frac{\delta L}{L(t)}\,\,n_2(L(t),t) \nonumber\\
&  &+\,2\,p_2\,\delta t\int_y ^{y_m(t)}n_1(z,t)\,\frac{dz}{z}\,+2\,p_3\,\delta t\int_{y_m(t)}^{L(t)}n_2(z,t)\,\frac{dz}{z} \nonumber\\
&  &+\,2\,p_4\,\delta t\int_{y_m(t)}^{L(t)}n_2(z,t)\,\ln\left(\frac{z}{\tilde y(t)}\right)\frac{dz}{z}
\end{eqnarray}

\noindent%
for $y < y_m(t)$ and 

\begin{eqnarray}  \label{flpeq4}
n_2(y,t+\delta t)& = &n_2(y,t)\;-\;p_3\,\delta t\,n_2(y,t)\;-p_4\,\delta t \,\ln\left(\frac{y}{\tilde y(t)}\right)n_2(y,t)  \nonumber\\
&  &+\:2\,p_1\,\frac{\delta L}{L(t)}\,n_2(L(t),t)\;+\;2\,p_3\,\delta t\int_{y}^{L(t)} n_2(z,t)\,\frac{dz}{z}  \nonumber\\
&  &+\:2\,p_4\,\delta t\int_y^{L(t)} n_2(z,t)\,\ln\left(\frac{z}{\tilde y(t)}\right)\frac{dz}{z} 
\end{eqnarray}

\noindent%
for $y > y_m(t)$. Here $n_1(y,t)$ and $n_2(y,t)$ are the densities of fragments of size $y$ at time $t$ which is smaller and larger than $y_m(t)$, respectively. The second term on the right hand side of (\ref{flpeq3}) and the second and third terms on the right hand side of (\ref{flpeq4}) represent the decrease in the number of particles of size $y$ from fragmentation into smaller particles, with probabilities $p_2$, $p_3$ and $p_4$, respectively. The next term on the right hand sides of (\ref{flpeq3},\ref{flpeq4}) describes the gain in the number of particles of size $y$ from the fragmentation of the largest fragment of size $L(t)$. The remaining terms on the right hand sides of (\ref{flpeq3},\ref{flpeq4}) are the increase in the number of particles of size $y$ from the fragmentation of all particles larger than $y$ and smaller than $L(t)$.

Now we define densities of fragments of length $y$ at time $t$, which are normalised to a positive constant value A for two regions of sizes smaller and larger than $y_m(t)$,

\begin{equation}  \label{flpeq5}
g_1(y,t) = A\,\frac{n_1(y,t)}{\displaystyle \int_0 ^{y_m(t)} n_1(z,t)\,dz\;+\;\displaystyle \int_{y_m(t)} ^{L(t)} n_2(z,t)\,dz}
\end{equation}	

\noindent%
for $y < y_m(t)$ and

\begin{equation}  \label{flpeq6}
g_2(y,t) = A\,\frac{n_2(y,t)}{\displaystyle \int_0 ^{y_m(t)} n_1(z,t)\,dz\;+\;\displaystyle \int_{y_m(t)} ^{L(t)} n_2(z,t)\,dz}
\end{equation}

\noindent%
for $y > y_m(t)$. In time $t\to t\,+\,\delta t$ as $L(t)\to L(t)-p_1\,\delta L$ and $y_m(t)\to y_m(t)-\delta y_m$ they yield

\begin{equation}  \label{flpeq7}
g_1(y,t+\delta t) = A\,\frac{n_1(y,t+\delta t)}{\displaystyle \int_0 ^{y_m(t)-\delta y_m(t)} n_1(z,t+\delta t)\,dz\;+\;\displaystyle \int_{y_m(t)-\delta y_m(t)} ^{L(t)-p_1\delta L}n_2(z,t+\delta t)\,dz}
\end{equation}		

\noindent%
for $y < y_m(t)$ and

\begin{equation}  \label{flpeq8}
g_2(y,t+\delta t) = A\,\frac{n_2(y,t+\delta t)}{\displaystyle \int_0 ^{y_m(t)-\delta y_m(t)} n_1(z,t+\delta t)\,dz\;+\;\displaystyle \int_{y_m(t)-\delta y_m(t)} ^{L(t)-p_1\delta L}n_2(z,t+\delta t)\,dz}
\end{equation}		

\noindent%
for $y > y_m(t)$. Inserting (\ref{flpeq3}), (\ref{flpeq4}) in (\ref{flpeq7}), (\ref{flpeq8}), respectively, gives us 

\begin{eqnarray}  \label{flpeq9}
\lefteqn{g_1(y,t+\delta t) = } \nonumber\\
&  & 2\,p_1\,\frac{\delta L}{L(t)}\,g_2(L(t),t)\,+\,2\,p_2\,\delta t\int_y ^{y_m(t)}g_1(z,t)\,\frac{dz}{z} \nonumber\\
&  & +\;2\,p_3\,\delta t \int_{y_m(t)}^{L(t)} g_2(z,t)\frac{dz}{z}\,+\,2\,p_4 \,\delta t \int_{y_m(t)}^{L(t)} g_2(z,t) \ln \left(\frac{z}{\tilde y(t)}\right)\frac{dz}{z} \nonumber\\
&  & +\;g_1(y,t)\Bigg[1-p_2\,\delta t -\,\frac{p_1}{A}\,g_2(L(t),t)\,\delta L\,-\,\frac{p_2}{A}\,\delta t\int_0 ^{y_m(t)}g_1(z,t)dz \nonumber\\
&  &-\;\frac{p_3}{A}\,\delta t\int_{y_m(t)}^{L(t)}g_2(z,t)dz\,-\,\frac{p_4}{A} \,\delta t \int_{y_m(t)}^{L(t)} g_2(z,t)\ln \left(\frac{z}{\tilde y(t)}\right) dz \Bigg]
\end{eqnarray}     		

\noindent%
for $y < y_m(t)$ and

\begin{eqnarray}  \label{flpeq10}
\lefteqn{g_2(y,t+\delta t) = } \nonumber\\
&  & 2\,p_1\,\frac{\delta L}{L(t)}\,g_2(L(t),t)\,+\,2\,p_3\,\delta t\int_y ^{L(t)}g_2(z,t)\,\frac{dz}{z} \nonumber\\
&  &+\;2\,p_4\,\delta t \int_y^{L(t)} g_2(z,t)\ln \left(\frac{z}{\tilde y(t)}\right) \frac{dz}{z}\,+\,g_2(y,t)\Bigg[1\,-\,p_3\,\delta t \nonumber\\
&  & -\;p_4 \,\delta t \,\ln \left(\frac{y}{\tilde y(t)}\right)-\:\frac{p_1}{A}\,g_2(L(t),t)\,\delta L- \frac{p_2}{A}\,\delta t\int_0 ^{y_m(t)}g_1(z,t)dz \nonumber\\
&  &-\;\frac{p_3}{A}\,\delta t\int_{y_m(t)}^{L(t)}g_2(z,t)dz\,-\frac{p_4}{A}\, \delta t \int_{y_m(t)}^{L(t)} g_2(z,t)\ln {\left(\frac{z}{\tilde y(t)}\right)} dz \Bigg] 
\end{eqnarray}

\noindent%
for $y > y_m(t)$. In order to solve (\ref{flpeq9}) and (\ref{flpeq10}) we introduce two functions $F_1(x,t)$ and $F_2(x,t)$ defined by

\begin{equation}  \label{flpeq11}
F_1(x,t) = L(t)\,g_1(xL(t),t)
\end{equation}

\noindent%
for $x < x_m$ and

\begin{equation}  \label{flpeq12}
F_2(x,t) = L(t)\,g_2(xL(t),t)
\end{equation}

\noindent%
for $x > x_m$. Here we have defined dimensionless and time independent variables $x = \displaystyle\frac{y(t)}{L(t)}$, $x_m = \displaystyle\frac{y_m(t)}{L(t)}$ and $\tilde x = \displaystyle\frac{\tilde y(t)}{L(t)}$, where $x$ changes in the range $[0,1]$. $x_m$ and $\tilde x$ are fixed values and restricted to the same range. Using (\ref{flpeq9}-\ref{flpeq12}) we can obtain two partial differential equations for $F_1(x,t)$ and $F_2(x,t)$ as

\begin{eqnarray}  \label{flpeq13}
\lefteqn{\frac{\partial F_1(x,t)}{\partial t} \;= }  \nonumber\\
&  &-\;\nu\,x\,\frac{\partial F_1(x,t)}{\partial x} + 2\,\nu \,F_2(1,t) +\,2\,p_2\int_x^{x_m} F_1(z,t)\,\frac{dz}{z}\,+ 2\,p_3\int_{x_m}^1 F_2(x,t)\,\frac{dx}{x}  \nonumber \\
&  & +\;2\, p_4\int_{x_m}^1 F_2(x,t)\ln \left(\frac{x}{\tilde x}\right)\frac{dx}{x}\,-\,F_1(x,t) \Biggm[\nu \,+\, p_2\, +\,\frac{\nu}{A}\, F_2(1,t)    \nonumber \\
&   & +\;\frac{p_2}{A}\int_0^{x_m}F_1(x,t)dx \,+\,\frac{p_3}{A}\int_{x_m}^1 F_2(x,t)dx \,+\,\frac{p_4}{A}\int_{x_m}^1 F_2(x,t)\ln \left(\frac{x}{\tilde x}\right)dx \Biggm]  \nonumber\\       
\end{eqnarray}

\noindent%
for $x < x_m$ and

\begin{eqnarray}  \label{flpeq14}
\lefteqn{\frac{\partial F_2(x,t)}{\partial t} = }  \nonumber\\
&  &-\;\nu\,x\,\frac{\partial F_2(x,t)}{\partial x} + 2\,\nu \,F_2(1,t) +\,2\,p_3\int_x^1 F_2(z,t)\,\frac{dz}{z} \nonumber \\
&  & +\;2\,p_4\int_x^1 F_2(z,t)\ln \left(\frac{z}{\tilde x}\right)\frac{dz}{z}-\;F_2(x,t) \Biggm[\nu\,+\, p_3  \nonumber\\  
&  &+\;p_4\ln \left(\frac{x}{\tilde x}\right)\,+\frac{\nu}{A}\, F_2(1,t)\,+ \frac{p_2}{A}\int_0^{x_m}F_1(x,t)dx     \nonumber \\
&   & +\;\frac{p_3}{A}\int_{x_m}^1 F_2(x,t)dx \,+\,\frac{p_4}{A}\int_{x_m}^1 F_2(x,t)\ln \left(\frac{x}{\tilde x}\right)dx \Biggm]         
\end{eqnarray}

\noindent%
for $x > x_m$. Here we have set $p_1\displaystyle\frac{\delta L}{L(t)\delta t} = \nu$ to be a fixed value, and assumed that $p_1 > 0$. $\nu$ is equal to the rate of fragmentation for the largest fragment, and determines the relationship between the real time and the length of the largest fragment at time $t$, $L(t)$. In the limit $t \to \infty$, we assume that $F_1(x,t)$ and $F_2(x,t)$ evolve to time-independent quantities,

\begin{equation}  \label{flpeq15}
F_1(x) = \lim\nolimits_{\,t\to \,\infty}\;F_1(x,t)
\end{equation}
for $x < x_m$ and

\begin{equation}  \label{flpeq16}
F_2(x) = \lim\nolimits_{\,t\to \,\infty}\;F_2(x,t)
\end{equation}

\noindent%
for $x > x_m$. So that, as $t \to \infty$ then $\displaystyle\frac{\partial F_1(x,t)}{\partial t}\to 0$ and $\displaystyle\frac{\partial F_2(x,t)}{\partial t}\to 0$. To solve (\ref{flpeq13},\ref{flpeq14}) in the long time limit using (\ref{flpeq15},\ref{flpeq16}), we anticipate solutions as

\begin{equation}  \label{flpeq17}
F_1(x) = B\,x^{-\,\alpha}
\end{equation}

\noindent%
for $x < x_m$ and 

\begin{equation}  \label{flpeq18}
F_2(x) = C\,\frac{\exp \left\{-\,\beta\,\left[\ln \left(\frac{x}{\tilde x}\right)\right]^2\right\}}{x}
\end{equation}

\noindent%
for $x > x_m$. Substituting (\ref{flpeq17},\ref{flpeq18}) in (\ref{flpeq13},\ref{flpeq14}) in the long time limit, using (\ref{flpeq15},\ref{flpeq16}) give us

\begin{equation}  \label{flpeq19}
\alpha\; =\; \frac{p_2}{p_3},
\end{equation}

\begin{equation}  \label{flpeq20}
\beta\; =\; \frac{p_4}{2\,p_3}
\end{equation}

\noindent%
and

\begin{equation}  \label{flpeq21}
\nu \; =\; p_3 .
\end{equation}

\noindent%
B and C in (\ref{flpeq17}),(\ref{flpeq18}), respectively, are constant values and can be obtained using (\ref{flpeq13},\ref{flpeq14}) in the long time limit and the continuity condition at $x_m$,

\begin{equation}  \label{flpeq22}
F_1(x_m)\; =\; F_2(x_m).
\end{equation}

\noindent%
These give 

{\small
\begin{equation}  \label{flpeq23}
\displaystyle C = \displaystyle\frac{\displaystyle A}{\displaystyle \left(\frac{p_3}{p_3-p_2}\right)\,\exp \left\{-\,\beta\left[\ln\left(\frac{x_m}{\tilde x}\right)\right]^2\right\}\,+\,\sqrt{\frac{2\pi p_3}{p_4}}\left[P\left(\sqrt{\frac{p_4}{p_3}}\,\ln\left(\frac{1}{\tilde x}\right)\right)\,-\,P\left(\sqrt{\frac{p_4}{p_3}}\,\ln \left(\frac{x_m}{\tilde x}\right)\right)\right]}
\end{equation}
}

\noindent%
and

\begin{equation}  \label{flpeq24}
B = C\,\displaystyle \frac{\displaystyle \exp \left\{-\,\frac{p_4}{2p_3}\left[\ln \left(\frac{x_m}{\tilde x}\right)\right]^2\right\}}{\displaystyle x_m^{\left(1\,-\,\frac{p_2}{p_3}\right)}},
\end{equation}

\noindent%
where the second and third terms in the denominator of (\ref{flpeq23}) are a proportion of the normal probability distribution function defined as

\begin{equation}  \label{flpeq25}
P(x) = \frac{1}{\sqrt{2\pi}}\int_{-\,\infty}^x e^{-\,t^2 /2}dt.
\end{equation}

\noindent%
We have also assumed that

\begin{equation}  \label{flpeq26}
0\,\le\alpha = \frac{p_2}{p_3} < 1.
\end{equation}

\noindent%
Solutions (\ref{flpeq17},\ref{flpeq18}) using (\ref{flpeq19},\ref{flpeq20},\ref{flpeq23}-\ref{flpeq26}) satisfy the normalisation relation

\begin{equation}  \label{flpeq27}
\int_0^{x_m} F_1(x)\,dx \,+\, \int_{x_m}^1 F_2(x)\,dx = A.
\end{equation}

\noindent%
If $A = 1$, the normalisation is equal to $1$. For a choice of $A$ as

{\small
\begin{equation}  \label{flpeq28}
A = \sqrt{\frac{p_4}{2\pi p_3}}\;\frac{p_3}{p_3\,-\,p_2}\exp \left\{-\,\beta\left[\ln \left(\frac{x_m}{\tilde x}\right)\right]^2\right\}\,+\,P\left(\sqrt{\frac{p_4}{p_3}}\, \ln\left(\frac{1}{\tilde x}\right)\right)\,-\,P\left(\sqrt{\frac{p_4}{p_3}}\, \ln \left(\frac{x_m}{\tilde x}\right)\right),
\end{equation}
}

\noindent%
the solutions (\ref{flpeq17},\ref{flpeq18}) using (\ref{flpeq19},\ref{flpeq20},\ref{flpeq23},\ref{flpeq24}) yield

\begin{equation}  \label{flpeq29}
F_1(x) = \sqrt{\frac{p_4}{2\pi p_3}}\;\displaystyle \frac{\displaystyle \exp \left\{-\,\frac{p_4}{2p_3}\left[\ln \left(\frac{x_m}{\tilde x}\right)\right]^2\right\}}{x_m^{\left(1\,-\,\frac{p_2}{p_3}\right)}}\;x^{-\,\frac{p_2}{p_3}}
\end{equation}

\noindent%
for $x < x_m$ and

\begin{equation}  \label{flpeq30}
F_2(x) = \sqrt{\frac{p_4}{2\pi p_3}}\:\displaystyle \frac{\displaystyle \exp \left\{-\,\frac{p_4}{2p_3}\left[\ln \left(\frac{x}{\tilde x}\right)\right]^2\right\}}{x}
\end{equation}

\noindent%
for $x > x_m$. We see that the stable distribution has a power law behaviour in the smaller region, $x < x_m$. For a log-normal distribution, the probability of finding a fragment of size between $x$ and $x + dx$ is given by $n(x)dx$, where

\begin{equation}  \label{flpeq31}
n(x) = \frac{\exp\left\{-\left[\ln\left(x/\tilde x \right)\right]^2 / 2\sigma ^2\right\}}{\left(2\pi \sigma ^2\right)^{1/2} x},
\end{equation}

\noindent%
where $\tilde x$ is the mean and $\sigma $ is the dispersion of the distribution. Comparing (\ref{flpeq30}) with (\ref{flpeq31}) shows that the stable distribution $F_2(x)$ in the larger region exhibits a log-normal behaviour with a mean and a dispersion of the distribution equal to $\tilde x$ and $\sigma = \displaystyle\sqrt{\frac{p_3}{p_4}}$, respectively. To have a positive value for $A$ in (\ref{flpeq28}) and therefore physically meaningful distributions, we require $p_3 > p_2$. This requirement is satisfied in (\ref{flpeq26}). 

Now we consider some special cases of the model.

\renewcommand{\labelenumi}{\arabic{enumi}. }

\begin{enumerate}

\item
There is no fragmentation for the fragments smaller than $y_m(t)$, i.e. $p_2 = 0$ and $p_1 + p_3 + p_4 = 1$. The stable distribution then from (\ref{flpeq29}) yields

\begin{equation}  \label{flpeq32}
F_1(x) = \sqrt{\frac{p_4}{2\pi p_3}}\;\,\displaystyle\frac{\exp \left\{\displaystyle -\,\frac{p_4}{2p_3}\left[\ln \left(\displaystyle\frac{x_m}{\tilde x}\right)\right]^2\right\}}{x_m}
\end{equation}

\noindent%
for $x < x_m$ and is the same as (\ref{flpeq30}) for $x > x_m$. We see that the distribution is a constant value in the smaller region and has a log-normal form in the larger region.

\item
As $y_m(t) = \tilde y(t)$ and therefore $x_m = \tilde x$, the distributions (\ref{flpeq29},\ref{flpeq30}) give

\begin{equation}  \label{flpeq33}
F_1(x) = \sqrt{\frac{p_4}{2\pi p_3}}\;x_m^{-\,\left(1\,-\,\frac{p_2}{p_3}\right)}\,x^{-\,\frac{p_2}{p_3}}
\end{equation}

\noindent%
for $x < x_m$ and 

\begin{equation} \label{flpeq34}
F_2(x) = \sqrt{\frac{p_4}{2\pi p_3}}\;\,\displaystyle\frac{\exp \left\{\displaystyle -\,\frac{p_4}{2p_3}\left[\ln \left(\displaystyle\frac{x}{x_m}\right)\right]^2\right\}}{x}
\end{equation}

\noindent%
for $x > x_m$. Again the distribution reveals a power law form in the smaller region and has a log-normal behaviour in the larger region.

\item
As $p_4 \to 0$ and therefore $\sigma ^2 = \displaystyle\frac{p_3}{p_4} \to \infty$, the distribution (\ref{flpeq30}) in the larger region yield $F_2(x) \sim x^{-\,1}$. On the other hand as $p_4 \to 0$ the rate of fragmentation (\ref{flpeq2}) for the fragments of sizes larger than $y_m(t)$ is only proportional to the inverse of the fragment size. Consequently, at each time step either the largest fragment breaks with probability $p_1$, or a fragment smaller than $y_m(t)$ is chosen for fragmentation with probability $p_2$, or a fragment larger than $y_m(t)$ is selected with probability $p_3$, where $p_1 + p_2 + p_3 = 1$. The stable distributions then exhibit a composite power law behaviour with power laws of exponents equal to $-\,\displaystyle\frac{\displaystyle p_2}{\nu}$ and $\displaystyle -\,\frac{p_3}{\nu}$ in the smaller and larger region, respectively. In our case since from (\ref{flpeq21}) we have $\nu = p_3$, the exponents reduce to $-\,\displaystyle\frac{p_2}{p_3}$ and $-1$ in the two regions. The power law distribution in the smaller region is similar to that from (\ref{flpeq17},\ref{flpeq19}) and in the larger region has an exponent of $-1$ as expected. The latter is consistent with one of interesting properties of the log-normal distribution where as the variance $\sigma ^2$ is pretty large, then the distribution can be approximated by a power law form with an exponent equal to $-1$, \cite{flpref18}.

\end{enumerate}

\section{Model B}
\label{flpsection3}
 
In this model again a time dependent transition size, $y_m(t)$, produces two regions of fragment sizes with different rates of fragmentation above and below the transition size. At each time step either the largest fragment in the system breaks with a fixed probability, $p_1$, or a fragment smaller than the transition size is chosen for the fragmentation with probability $p_2$ and a rate proportional to the inverse of fragment size, or another fragment larger than $y_m(t)$ is selected with probability $(1 - p_1 - p_2)$ and a rate proportional to $\displaystyle\frac{\delta t}{y(t)}\left[1+\lambda \ln\left(\frac{y(t)}{\tilde y(t)}\right)\right]$. Here $y(t)$ is the size of fragment at time $t$, $\lambda$ is a positive and constant value and $\tilde y(t)$ is a size for which the logarithmic term is zero at time $t$. In this model we assume that $y_m(t) < \tilde y(t)$. For a fragment of size $y > \tilde y(t)$ the logarithmic term in the rate of fragmentation is always positive, but a fragment of size $y_m(t) < y < \tilde y(t)$ gives a negative value for the logarithmic term. In order to have a positive rate of fragmentation for all particles larger than $y_m(t)$, we require $\left[1+\lambda \ln\left(\displaystyle\frac{y(t)}{\tilde y(t)}\right)\right] > 0$ which gives $y(t) > \tilde y(t)\,e^{-\,1/\lambda}$. If we choose $y_m(t)$ to be  

\begin{equation}  \label{flpeq35}
y_m(t) = \tilde y(t)\,e^{-\,1/\lambda},
\end{equation}

\noindent%
then the rate of fragmentation for all fragments of sizes larger than $y_m(t)$ is positive.

Consequently, the rate of fragmentation for this model is given by

\begin{eqnarray} \label{flpeq36}
\delta t \,R(y,z)& \,= &\,p_1 \,\frac{\delta L}{L(t)} \,\delta (y\,+\,z-L(t)) \,+ \,p_2 \,\,\frac{\delta t}{y+z} \,\theta (y_m(t)-(y+z)) \nonumber\\
&  &+\;(1-p_1-p_2)\,\frac{\delta t}{y+z}\left[1\,+\,\lambda\,\ln\left(\frac{y+z}{\tilde y(t)}\right)\right]\theta ((y+z)-y_m(t)), \nonumber\\
\end{eqnarray}	    

\noindent%
where $y_m(t)$ is defined as (\ref{flpeq35}). The first two terms on the right hand side of (\ref{flpeq36}) are the same as those in (\ref{flpeq2}) and have the same definitions as before. The last term represents the rate of fragmentation for the fragments of sizes larger than $y_m(t)$ in time $(t, t + \delta t)$. Using (\ref{flpeq1},\ref{flpeq36}) for the both size regions give the kinetic equations 

\begin{eqnarray}   \label{flpeq37}
\lefteqn{n_1(y,t+\delta t) = }  \nonumber\\
&  &n_1(y,t)\;-\;p_2\,\delta t \,n_1(y,t)\;+\;2\,p_1 \,\frac{\delta L}{L(t)}\,n_2(L(t),t) \nonumber\\
&  &+\:2\,p_2\,\delta t\int_y ^{y_m(t)}n_1(z,t)\,\frac{dz}{z}\,+2\,(1-p_1-p_2)\,\delta t\int_{y_m(t)}^{L(t)}n_2(z,t)\,\frac{dz}{z} \nonumber\\
&  &+\:2\,\lambda\,(1-p_1-p_2)\,\delta t\int_{y_m(t)}^{L(t)}n_2(z,t)\,\ln\left(\frac{z}{\tilde y(t)}\right)\frac{dz}{z}
\end{eqnarray}		                                                              	
\noindent%
for $y < y_m(t)$ and

\begin{eqnarray}  \label{flpeq38}
\lefteqn{n_2(y,t+\delta t) = }  \nonumber\\
& &n_2(y,t)-(1-p_1-p_2)\,\delta t\,n_2(y,t)-\,\lambda\,(1-p_1-p_2)\,\delta t \ln\left(\frac{y}{\tilde y(t)}\right)n_2(y,t)  \nonumber\\
&  &+\:2\,p_1\,\frac{\delta L}{L(t)}\,n_2(L(t),t)\:+\:2\,(1-p_1-p_2)\,\delta t\int_{y}^{L(t)} n_2(z,t)\,\frac{dz}{z}  \nonumber\\
&  &+\:2\,\lambda\,(1-p_1-p_2)\,\delta t\int_y^{L(t)} n_2(z,t)\,\ln\left(\frac{z}{\tilde y(t)}\right)\frac{dz}{z} 
\end{eqnarray}

\noindent%
for $y > y_m(t)$. Following the same procedure as that used in the model A, we obtain

\begin{eqnarray}  \label{flpeq39}
\lefteqn{\frac{\partial F_1(x,t)}{\partial t} = }  \nonumber\\
&  &-\;\nu\,x\,\frac{\partial F_1(x,t)}{\partial x} + 2\,\nu \,F_2(1,t) +\,2\,p_2\int_x^{x_m} F_1(z,t)\,\frac{dz}{z} \nonumber \\
&  &+\; 2\,(1-p_1-p_2)\int_{x_m}^1 F_2(x,t)\frac{dx}{x} +\,2\,\lambda\,(1-p_1-p_2)\int_{x_m}^1 F_2(x,t)\ln \left(\frac{x}{\tilde x}\right)\frac{dx}{x} \nonumber \\
&   & -\;F_1(x,t) \Bigg[\nu\,+\, p_2 \,+\,\frac{\nu}{A}\, F_2(1,t)  \,+\,\frac{p_2}{A}\int_0^{x_m}F_1(x,t)dx  \nonumber\\
&  &+\;\frac{(1-p_1-p_2)}{A}\int_{x_m}^1 F_2(x,t)dx\, +\,\frac{\lambda\,(1-p_1-p_2)}{A}\int_{x_m}^1 F_2(x,t)\ln \left(\frac{x}{\tilde x}\right)dx \Bigg] \nonumber\\        
\end{eqnarray}

\noindent%
for $x < x_m$ and

\begin{eqnarray}  \label{flpeq40}
\lefteqn{\frac{\partial F_2(x,t)}{\partial t} = } \nonumber\\
&  &-\;\nu\,x\,\frac{\partial F_2(x,t)}{\partial x} + 2\,\nu \,F_2(1,t) +\,2\,(1-p_1-p_2)\int_x^1 F_2(z,t)\,\frac{dz}{z} \nonumber \\
&  &+\: 2\,\lambda\,(1-p_1-p_2)\int_x^1 F_2(z,t)\ln \left(\frac{z}{\tilde x}\right)\frac{dz}{z}-\,F_2(x,t) \Bigg[\nu +(1-p_1-p_2)      \nonumber \\
&   &+\:\lambda\,(1-p_1-p_2)\,\ln \left(\frac{x}{\tilde x}\right)\, +\frac{\nu}{A}\, F_2(1,t)\,+ \frac{p_2}{A}\int_0^{x_m}F_1(x,t)dx   \nonumber\\
& &+\:\frac{(1-p_1-p_2)}{A}\int_{x_m}^1 F_2(x,t)dx  +\,\frac{\lambda\,(1-p_1-p_2)}{A}\int_{x_m}^1 F_2(x,t)\ln \left(\frac{x}{\tilde x}\right)dx \Bigg]  \nonumber\\       
\end{eqnarray}

\noindent%
for $x > x_m$. Here $x$, $x_m$, $\tilde x$ and $\nu$ have the same definitions as before and from (\ref{flpeq35}) we get
 
\begin{equation} \label{flpeq41}
x_m = \tilde x \,e^{-\,1/\lambda}.
\end{equation}

\noindent%
In the long time limit as $t \to \infty$, we assume that $F_1(x,t)$ and $F_2(x,t)$ evolve to the same time-independent variables as (\ref{flpeq15}) and (\ref{flpeq16}), respectively.
In order to solve (\ref{flpeq39},\ref{flpeq40}) in this limit we anticipate solutions as

\begin{equation}  \label{flpeq42}
F_1(x) = G\,x^{-\,\gamma}
\end{equation}

\noindent%
for $x < x_m$ and 

\begin{equation}  \label{flpeq43}
F_2(x) = H\:\frac{\exp \left\{-\,\eta\left[\ln\left(\frac{x}{\tilde x}\right)\right]^2\right\}}{x}
\end{equation}

\noindent%
for $x > x_m$. Substituting (\ref{flpeq42},\ref{flpeq43}) in (\ref{flpeq39},\ref{flpeq40}) in the long time limit, gives us

\begin{equation}  \label{flpeq44}
\gamma\; = \;\frac{p_2}{1-p_1-p_2},
\end{equation}

\begin{equation}  \label{flpeq45}
\eta \; =\; \frac{\lambda}{2}
\end{equation}
and

\begin{equation} \label{flpeq46}
\nu\; = \;1-p_1-p_2.
\end{equation}

\noindent%
G and H in (\ref{flpeq42},\ref{flpeq43}) are constant values and can be obtained using (\ref{flpeq39},\ref{flpeq40}) in the long time limit and the continuity condition at $x_m$, (\ref{flpeq22}). These give

\begin{equation} \label{flpeq47}
H = \displaystyle\frac{A}{\displaystyle\frac{1-p_1-p_2}{1-p_1-2p_2}\,\,\exp \left(\displaystyle -\,\frac{1}{2\lambda}\right) + \sqrt{\frac{2\pi}{\lambda}}\left[P\left(\sqrt{\lambda}\ln \left(\frac{1}{\tilde x}\right)\right) - P\left(-\,\frac{1}{\sqrt{\lambda}}\right)\right]}
\end{equation}
and

\begin{equation}  \label{flpeq48}
G = H\,\displaystyle\frac{\displaystyle\exp\left(\displaystyle -\,\frac{1}{2\lambda}\right)}{\displaystyle \left(\tilde x\,e^{-\,1/\lambda}\right)^\frac{1-p_1-2p_2}{1-p_1-p_2}},
\end{equation}
where we have assumed that

\begin{equation}  \label{flpeq49}
0\, \le \gamma = \frac{p_2}{1-p_1-p_2} < 1.
\end{equation}

\noindent%
The second and third terms in the denominator of (\ref{flpeq47}) are a proportion of the normal probability distribution function with the same definition as (\ref{flpeq25}). Solutions (\ref{flpeq42},\ref{flpeq43}) using (\ref{flpeq44}-\ref{flpeq49}) satisfy the normalisation relation (\ref{flpeq27}). If $A = 1$, the normalisation is equal to $1$. For a choice of $A$ as

\begin{equation}  \label{flpeq50}
A = \sqrt{\frac{\lambda}{2\pi}} \;\,\frac{1-p_1-p_2}{1-p_1-2p_2}\,\exp\left(-\,\frac{1}{2\lambda}\right) + P\left(\sqrt{\lambda}\ln\left(\frac{1}{\tilde x}\right)\right) - P\left(-\,\frac{1}{\sqrt{\lambda}}\right),
\end{equation} 

\noindent%
the solutions (\ref{flpeq42},\ref{flpeq43}) using (\ref{flpeq44},\ref{flpeq45},\ref{flpeq47},\ref{flpeq48}) yield

\begin{equation}  \label{flpeq51}
F_1(x) = \sqrt{\frac{\lambda}{2\pi}}\;\displaystyle\frac{\displaystyle\exp\left(\displaystyle -\,\frac{1}{2\lambda}\right)}{\displaystyle \left(\tilde x\,e^{-\,1/\lambda}\right)^\frac{1-p_1-2p_2}{1-p_1-p_2}}\;x^{-\,\frac{p_2}{1-p_1-p_2}}
\end{equation}

\noindent%
for $x < \tilde x\,e^{-\,1/\lambda}$ and

\begin{equation}  \label{flpeq52}
F_2(x) = \sqrt{\frac{\lambda}{2\pi}}\;\,\displaystyle\frac{\exp\left\{\displaystyle -\,\frac{\lambda}{2}\left[\ln\left(\displaystyle\frac{x}{\tilde x}\right)\right]^2\right\}}{x}
\end{equation}

\noindent%
for $x > \tilde x\, e^{-\,1/\lambda}$. We see that the stable distribution has a power law behaviour in the smaller region. Comparing (\ref{flpeq52}) with (\ref{flpeq31}) exhibits that the stable distribution $F_2(x)$ in the larger region has a log-normal form with a mean of $\tilde x$ and a dispersion of the distribution equal to

\begin{equation}  \label{flpeq53}
\sigma = \frac{1}{\sqrt{\lambda}}.
\end{equation}

\noindent%
It is shown that the log-normal function (\ref{flpeq52}) has a maximum at $x = \tilde x\,e^{-\,1/\lambda}$ which we have assumed to be equal to $x_m$. This shows that in this model the dimensionless transition size $x_m$ is chosen to be equal to the value for which the log-normal distribution in the larger region is maximum. To have a positive value for $A$ in (\ref{flpeq50}) and therefore physically meaningful distributions, it is required that $(1-p_1-p_2) > p_2$. This inequality is satisfied from (\ref{flpeq49}).

Now we consider some special cases of the model B.

\renewcommand{\labelenumi}{\arabic{enumi}. }

\begin{enumerate}

\item
When there is no fragmentation for the fragments smaller than $y_m(t)$, i.e. $p_2 = 0$, then the stable distribution from (\ref{flpeq51}) yields

\begin{equation}  \label{flpeq54}
F_1(x) = \sqrt{\frac{\lambda}{2\pi}}\;\frac{1}{\tilde x}\;e^{1/{2\lambda}}
\end{equation}

\noindent%
for $x < x_m$ and is the same as (\ref{flpeq52}) for $x > x_m$. The distribution is a constant value in the smaller region and has a log-normal behaviour in the larger region.

\item
As $\lambda \to 0$, and therefore from (\ref{flpeq53}) $\sigma ^2 \to \infty$, the situation is then similar to the case $3$ of model A and the log-normal distribution (\ref{flpeq52}) in the larger region can be approximated by a power law form with an exponent equal to $-1$, $F_2(x) \sim x^{-\,1}$. On the other hand at this limit the rate of fragmentation (\ref{flpeq36}) gives two different probabilities in the two different size regions with a rate proportional to the inverse of the fragment size in each region. The stable distribution then reveal a composite power law with power laws of exponents of $\displaystyle -\,\frac{p_2}{\nu}$ and $\displaystyle -\,\frac{1-p_1-p_2}{\nu}$ in the smaller and larger size regions, respectively. Since in our case from (\ref{flpeq46}) we have $\nu = (1-p_1-p_2)$, the exponents yield $\displaystyle -\,\frac{p_2}{1-p_1-p_2}$ and $-1$, respectively. The result in the smaller size region is equivalent to that from (\ref{flpeq42},\ref{flpeq44}). In the larger size region, as expected, we get a power law with an exponent of $-1$ which is consistent with the property of a log-normal distribution with a very large variance.
 
\end{enumerate}

We see that all the results for the model B can be recovered from the model A by substituting $p_3$ by $(1-p_1-p_2)$, $p_4$ by $\lambda\,(1-p_1-p_2)$ and using equation (\ref{flpeq35}).

\section{Conclusions}
\label{flpsection4}

In this paper we have studied two models of binary fragmentation. In the models we introduced a time dependent transition size, $y_m(t)$, which produced two regions of fragment sizes with different rates of fragmentation above and below the transition size. The rate of fragmentation at each time step is fixed for the largest fragments and is proportional to the inverse of fragment size for the fragments smaller than the transition size. We considered a rate of fragmentation combined of two terms, one proportional to the inverse of the fragment size and the other proportional to a logarithmic function of the fragment size, for the fragments larger than the transition size at each time step. For a size equal to $\tilde y(t)$ at time $t$, the logarithmic function is zero. In model A we assumed that the transition size, $y_m(t)$ is any arbitrary value for which $y_m(t) \geq \tilde y(t)$, whereas in the model B we chose a value as (\ref{flpeq35}) for the transition size. The models were then solved exactly in the long time limit to reveal stable time-invariant solutions for the distributions. The results of these distributions for both models A and B exhibited a power law form with an exponent between $-1$ and zero, in the smaller region. For a special choice of normalisation, the distributions exhibited log-normal behaviours in the larger regions. In this region they revealed one of interesting properties of a log-normal distribution for which as the variance of the log-normal is pretty large, then the distribution is approximated by a power law form with an exponent equal to $-1$. Special cases of the models with no fragmentation for the smaller fragments were also examined. The stable distributions then exhibited a constant value and a log-normal form in the smaller and larger regions, respectively.
 
Now we investigate some experimental work on the fragmentation involving log-normal distributions and the transition of the distributions from a log-normal form to a power law one and vice versa. In \cite{flpref20} the experimental results of fragment size and mass distributions for the long thin glass rods with fixed lengths of $1500mm$ and diameters of $2mm$ were reported. The rods were dropped horizontally onto a flat hard floor from different heights. At lower falling heights (about $1m$ drop), the distributions exhibited a log-normal form for larger fragments and a power law form for smaller ones. The crossover was seen to be at length scales around the rod diameter. This is due to the fact that fragments smaller than the rod diameter, undergo a three-dimensional fracture, whereas the larger fragments are produced by one-dimensional breaking. For a falling height of $1.20m$ the cumulative number of fragments on their size were plotted. The data was fitted to a curve calculated by assuming that the size distribution could be described by a log-normal distribution. The fitting gave values of $\tilde l = 30mm$ and $\sigma = 0.55$ for the mean and the dispersion of the log-normal distribution. The fitting is excellent for the fragments larger than $l_c = 7mm$ which is a length scale around the rod diameter. The size distribution of fragments smaller than $l_c$ seemed to have a power law form. As the falling height was increased, the cumulative number variation for larger fragments started to show a power law dependence on their size and mass. In \cite{flpref19} experiments in rupture of mercury droplets were performed. In the experiment at a height of $h$, the mercury droplets of about $2.0mm$ radius were fallen directly on a glass Petri dish. The results of the cumulative number of the drops versus the drop diameter were plotted. For small falling heights a log-normal behaviour was exhibited. The distribution showed a clear transition from a log-normal form to a scaled one as the falling height $h$ was increased. 

Our models in the long time limit can be applied to the experimental systems when the fragmentation process is over and the system reaches a stable and time-independent phase. In the models we assumed that the fragments have only one dimension, whereas in real physical systems the fragments have three dimensions. We can therefore only apply our models to the systems for which the fractures occur in one dimension. This assumption seems physically reasonable for a long thin glass rod \cite{flpref20} for which most of the fractures occur in the length of the rod and the cross-section remains almost fixed. We can also consider this assumption for the mercury droplets \cite{flpref19} for which at each time step only the radius of the droplet changes and the shape of the fragments remains always spherical.

Now we discuss the similarities between the results from our models and those from the experiments. The experimental results for the size distribution of long thin glass rods at low falling heights exhibited a power law and log-normal behaviour in the smaller and larger size regions, respectively \cite{flpref20}. We got the same behaviour for the distributions in both our models. The data from the experiments for the falling height of $1.20m$ revealed that $l_c < \tilde l$, where $l_c$ is the transition size and $\tilde l$ is the mean of the log-normal distribution. In the model B of our work also we have $x_m < \tilde x$. We therefore expect model B of our work to describe the experimental results. In model B, the dispersion of the log-normal distribution can be obtained from (\ref{flpeq53}) and the mean and the transition values, $\tilde x$ and $x_m$, respectively, are any arbitrary values which satisfy the relation (\ref{flpeq41}). Then, by appropriate choices of $\tilde x$ and $\lambda$ in the model B of our work we can match the mean and the dispersion of the log-normal distribution in the experimental results. Because of (\ref{flpeq41},\ref{flpeq53}) we should expect a relation as $l_c = \tilde l\,e^{-\,\sigma ^2}$. The values of $l_c$ and $\tilde l$ from experimental result do not satisfy this relation. As a result, although model B of our work does not match the exact relation between $l_c$ and $\tilde l$ from experimental result, but it can give an explanation for the transition of distribution from a power law to a log-normal form on the shock fragmentation of long thin glass rods \cite{flpref20}. Also in the smaller region of the model B, the appropriate choices of $p_1$ and $p_2$ can match the exponent of power law equal to $\displaystyle -\,\frac{\displaystyle p_2}{\displaystyle (1-p_1-p_2)}$ from (\ref{flpeq42},\ref{flpeq44}), with the corresponding exponent of power law distribution from the experimental results. This exponent needs to satisfy the condition (\ref{flpeq49}).
For high falling heights a transition of the distribution from log-normal to scaling law was observed in the experiments for the larger rod fragments \cite{flpref20} and for the mercury droplets \cite{flpref19}. At high heights therefore the distributions from the experiments exhibit a composite and a single power law form for the rod fragments and mercury droplets, respectively. The models in this paper suggest a power law and a log-normal behaviour for the distribution of the smaller and larger fragments, respectively. To explain composite and single power law distributions, we can however use the models in \cite{flpref10}.  

To conclude we introduced statistical models with a fixed probability for the breaking the largest particle in the system. All other particles of sizes smaller than the largest size, were divided into two different regions with different rates of fragmentation in each region. In the models therefore the largest fragments are broken with very large probability. This fact together with the form of the fragmentation rate as a function of the fragment size in each region, give a distribution of power law or log-normal form in that region. Our models therefore give a theoretical explanation for the transition of distribution from the scaling to log-normal behaviour, which had been seen before in the experiments.

\vskip 1cm
\noindent%
{\large \bf Acknowledgement}

\vskip 0.5cm
\noindent%
ZT is grateful to the Soudavar Scholarship Fund for the award of a scholarship.

\end{document}